# Short range magnetic order and effective suppression of superconductivity by manganese-doping in LaFe$_{1-x}$Mn$_x$AsO$_{1-y}$F$_y$

Rainer Frankovsky[1], Hubertus Luetkens[2], Frank Tambornino[1], Alexey Marchuk[1], Gwendolyne Pascua[2], Alex Amato[2], Hans-Henning Klauss[3] and Dirk Johrendt[1]*

[1] Department Chemie, Ludwig-Maximilians Universität München, 81377 München, Germany.

[2] Laboratory for Muon-Spin Spectroscopy, Paul Scherrer Institut, CH-5232 Villigen PSI, Switzerland.

[3] Institut für Festkörperphysik, TU Dresden, 01069 Dresden, Germany.

*Corresponding Author
Dirk Johrendt
Department Chemie
Ludwig-Maximilians Universität München
Butenandtsr. 5-13 (Haus D)
81377 München, Germany
PHONE: +49 89 2180 77430
FAX: +49 89 2180 77431
E-mail: johrendt@lmu.de

**Abstract**

We present a study of the structural and physical properties of directly hole doped LaFe$_{1-x}$Mn$_x$AsO ($x$ = 0.0-0.2) and the influence of charge compensation / electron-doping by additional F doping in LaFe$_{0.9}$Mn$_{0.1}$AsO$_{1-y}$F$_y$ ($y$ = 0.1-0.5). High quality polycrystalline samples were prepared using a solid state metathesis reaction. The unit cell increases upon Mn doping, but decreases again when additional F is inserted. The semiconducting character of LaFe$_{1-x}$Mn$_x$AsO decreases with additional F doping. Muon spin relaxation (µSR) measurements reveal short range magnetic order in LaFe$_{1-x}$Mn$_x$AsO and a suppression of magnetism by additional electron-doping with fluoride in LaFe$_{0.9}$Mn$_{0.1}$AsO$_{1-y}$F$_y$. Superconductivity remains absent even though the electronic preconditions are fulfilled in electron-doped LaFe$_{0.9}$Mn$_{0.1}$AsO$_{1-y}$F$_y$ at $x$ > 0.1, which is suggestive of effective pair breaking by Mn in this system.

## I. INTRODUCTION

Charge-doping suppresses structural transitions from tetragonal to orthorhombic symmetry and SDW antiferromagnetism in stoichiometric parent compounds of 1111- and 122-type iron arsenides like LaFeAsO or BaFe$_2$As$_2$, and superconductivity is induced in the proximity of magnetism.[1,2] In stark contrast to the cuprates, also substitution of the iron atoms by transition metals with the same or higher number of valence electrons induces superconductivity, for example in Ba(Fe$_{1-x}$Co$_x$)$_2$As$_2$.[3] However, whether transition metal substitution leads to charge doping or acts as scattering centers is still under debate.[4-6]

Even less understood is the fact that hole-doping by transition metal substitution (Cr, Mn) has on no account induced superconductivity so far.[7-10]

Substitution of Fe by Cr or Mn in Ba(Fe$_{1-x}$TM$_x$)$_2$As$_2$ leads to a suppression of the structural transition for $x$ ≥ 0.335 and $x$ > 0.10 respectively.[8-11] Even though the transition is absent for highly doped samples, a magnetic phase develops albeit with a magnetic structure different from the low doped samples.[9,12] The antiferromagnetic ordering changes from stripe-like (SDW) to Néel-type (G-type) in Cr substituted compounds, which is consistent with the absence of the orthorhombic phase. A co-existence of both types of AFM ordering is reported for $x$ = 0.305 and $x$ = 0.335, although the structural transition is already suppressed for the latter concentration.[9] This is unusual, since the stripe-like SDW is believed to be coupled to the orthorhombic transition by magnetoelastic effects.[13-15] For Ba(Fe$_{1-x}$Mn$_x$)$_2$As$_2$ a more complicated behavior is observed. In first reports, magnetic ordering with a propagation vector (½ ½ 1) (stripe-like or SDW AFM) was observed in the absence of the orthorhombic distortion ($x > 0.10$) but no traces of Néel type fluctuations or ordering have been found.[10] However, recent neutron diffraction studies found co-existence of long-ranged stripe like antiferromagnetic ordering and purely dynamic short-range Néel type spin fluctuations introduced by Mn in Ba(Fe$_{0.925}$Mn$_{0.075}$)$_2$As$_2$.[12] Whether these short-range fluctuations are also present in the samples with higher Mn concentrations has not been studied yet. NMR measurements of Ba(Fe$_{1-x}$Mn$_x$)$_2$As$_2$ revealed localized Mn moments, which couple to the conduction electrons and induce a staggered spin polarization within the Fe-layer.[6] It was proposed that spin fluctuations (Néel-type) which arise from these local Mn moments could be disruptive for superconductivity.[6, 12] Actually, very recent theoretical calculations confirmed the suppression of superconductivity by Néel type fluctuations in the iron pnictides.[16] Meanwhile there is growing evidence for local Mn moments leading to a different type of (short range) magnetic fluctuations / ordering, which competes with the long range ordering developed by the Fe lattice.

In contrast to this, very little is known about manganese substitution in 1111 compounds. Substitution of Fe by Mn in CaFeAsF and LaFeAsO changes the resistivity behavior from metallic to semiconducting.[7, 17] Because this is already observed for very small Mn concentrations one may argue that Mn mainly acts as a scattering center in these compounds[17]. In LaFe$_{1-x}$Mn$_x$AsO the structural distortion seems to be suppressed at $x > 0.06$ according to conductivity and thermoelectric power measurements.[7] However, information about the magnetic behavior of the 1111-compounds upon Mn substitution is still lacking.

It is widely believed that in FeAs superconductors certain structural preconditions like interatomic distances, the pnictogen layer height or the distance between the iron-arsenide layers have to be fulfilled to induce superconductivity or to reach high critical temperatures (for an overview see Ref. 18). One of the most noticeable correlations of $T_c$ to a structural parameter is found in the As–Fe–As angles of the FeAs tetrahedra. Lee *et al.* have collected structural data from many iron based superconductors and found that the highest $T_c$s appear in systems were the angles are close to the value of 109.47°, suggesting that the potential for high critical temperatures is biggest for regular tetrahedral.[19] It has been suggested that the angle is not only determined by the different atom sizes but that the electron count plays an important role as well.[20] Although the experimental data indicate that an ideal tetrahedral angle seems to be crucial for high $T_c$s, the most recent investigations about the interplay between doping and structural changes in doped BaFe$_2$As$_2$ have shown that charge modifications play the mayor role for the suppression of magnetism and the emergence of superconductivity. Zinth *et al.* have demonstrated that charge compensation in Ba$_{1-x}$K$_x$(Fe$_{0.93}$Co$_{0.07}$)$_2$As$_2$ ($x \approx 0.14$) recovers the magnetic and structural transitions of the parent compound and superconductivity re-emerges for lower (electron-doped) as well as for higher (hole-doped) potassium concentrations.[21] This shows how the physical properties can be controlled by modifying the charge balance in BaFe$_2$As$_2$.

To expand the knowledge about the influence of direct hole-doping on the structural, electronic and magnetic properties of 1111-type iron arsenides, we investigated the series LaFe$_{1-x}$Mn$_x$AsO ($x$ = 0.0 - 0.2) and draw a comparison with directly hole-doped 122 compounds. Furthermore the influence of charge compensation by additional electrons due to additional F doping in LaFe$_{0.1}$Mn$_{0.1}$AsO$_{0.9}$F$_{0.1}$ as well as the formally electron doped series LaFe$_{0.9}$Mn$_{0.1}$AsO$_{1-y}$F$_y$ ($y$ = 0.2-0.5) are presented.

## II. EXPERIMENTAL METHODS

FeAs was first synthesized by heating a stoichiometric mixture of the elements (pieces, purity > 99.8 %) to 773-973 K in sealed alumina crucibles and combined with stoichiometric amounts of Na (ingots, 99.8 %) to synthesize NaFeAs by heating to 1023 K in an niobium crucible sealed in a silica ampoule[22]. In contrast to NaFeAs, NaMnAs is better prepared from the elements than the corresponding binary metal arsenide [23]. The elements were heated at slow rates between 623 K and 873 K (10 h) and reacted at 1123 K for 48 h, followed by furnace cooling. LaOCl was synthesized by heating La$_2$O$_3$ (powder, 99.999 %) and NH$_4$Cl (powder, 99.5 %), in a molar ratio of 1 : 2.1, to 773-1173 K in a dynamic nitrogen atmosphere.[24] The synthesis of the LaFe$_{1-x}$Mn$_x$AsO$_{1-y}$F$_y$ samples was done by heating stoichiometric amounts of LaOCl, NaFeAs, NaMnAs, LaF$_3$ (powder, 99.99 %) and Na (ingots, 99.8 %) according to the solid state metathesis reaction (see Ref. 25 for details):

(1-$x$) NaFeAs + $x$ NaMnAs + (1-$y$) LaOCl + $y$ LaF$_3$ + $x$ Na
→ LaFe$_{1-x}$Mn$_x$AsO$_{1-y}$F$_y$ + (1-$y$) NaCl + 2$y$ NaF ($x$ = 0.05 – 0.20; $y$ = 0.0 – 0.5 with $x$ = 0.1)

The precursors were well homogenized, filled in alumina crucibles, welded in Niobium tubes enclosed by silica tubes. All crucibles and tubes contained a purified argon atmosphere. The reaction mixtures were heated to 1023 K for 48 h and 1223 K for 96 h, followed by cooling to room temperature with 300 K/h. The concomitantly formed salts NaCl and NaF were removed by washing the obtained mixture with water (3 times) and ethanol, followed by drying the product under high vacuum. Sample preparation except the synthesis of LaOCl and FeAs was performed in a glove box under an atmosphere of argon (O$_2$ and H$_2$O < 1 ppm).

Phase purity were checked by x-ray powder diffraction with Cu-K$_{\alpha 1}$ radiation (HUBER G670 Guinier imaging plate diffractometer) and Rietveld refinements using the TOPAS program package [26]. EDX measurements confirmed the effective Mn content of the series LaFe$_{1-x}$Mn$_x$AsO. The magnetic measurements were done with a Quantum Design MPMS XL5 SQUID magnetometer. The electric resistivity was measured from 12K to 300K using the four probe method. In order to detect even very weak or short range ordered magnetism, muon spin relaxation (µSR) experiments were performed at the πM3 beam line at the Paul-Scherrer-Institut (Switzerland).

## III. RESULTS AND DISCUSSION

The metathesis reaction yielded single phase samples of LaFe$_{1-x}$Mn$_x$AsO. With increasing Mn concentration the cell parameters and the volume increase almost linearly, indicating that Mn is successfully inserted. This, together with the sample purity, indicates that the effective Mn contents are very close to the nominal ones. Parallel to the unit cell expansion, the Fe–As and metal–metal distances increase. The twofold As–Fe–As angle $\varepsilon_2$ of the FeAs tetrahedra becomes more regular, decreasing from 113.1(2)° ($x = 0$) to 111.8(2)° ($x = 0.20$). The structural changes upon Mn-doping are therefore consistent with the ones reported by Bérardan *et al.* ($x = 0$-0.1),[7] but the reported cell parameters of the corresponding compositions are much larger than those we find. The effective doping levels have been confirmed by at least 5 EDX measurements of each sample and we obtained averaged compositions showing deviations of less than 1 % of the nominal compositions.

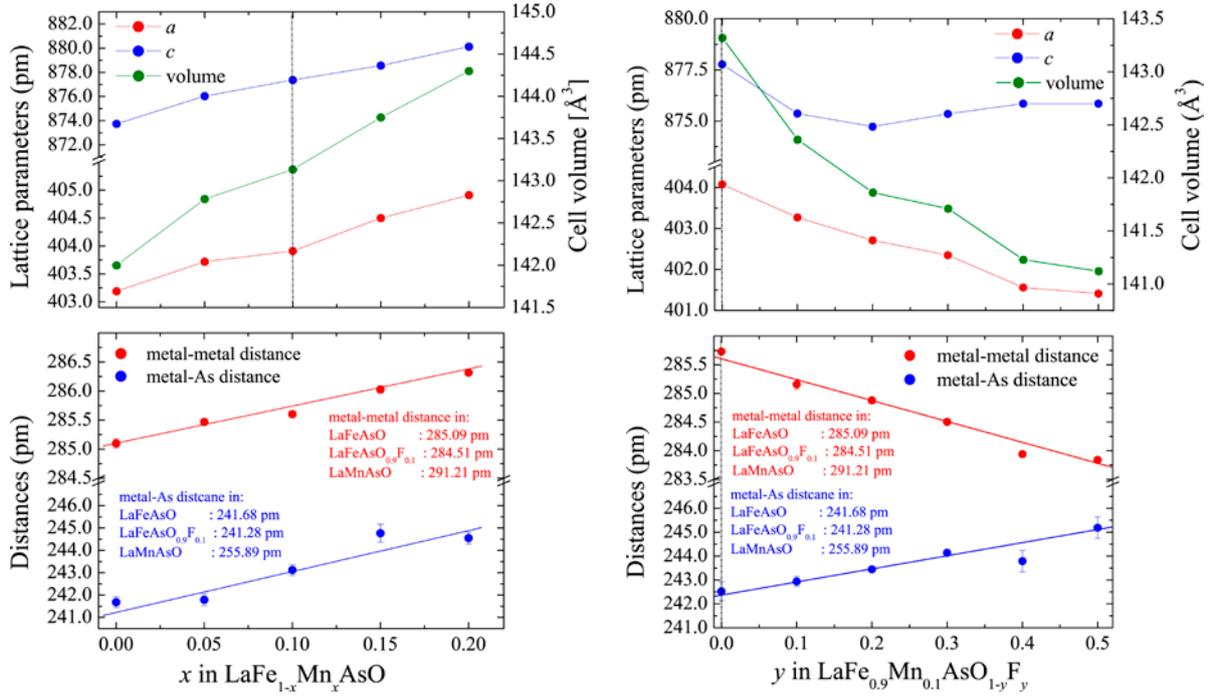

FIG. 1. Changes of the unit cell parameters (top) and interatomic distances (bottom) with increasing Mn fraction (left side) and further F doping in LaFe$_{0.9}$Mn$_{0.1}$AsO (right side). The dotted black line marks the Mn concentration held constant for the LaFe$_{0.9}$Mn$_{0.1}$AsO$_{1-y}$F$_y$ series. Some interatomic distances of similar 1111 compounds are listed for comparison (own synthesized compounds). Error bars of the cell parameters are smaller than the symbol size.

Samples of the series LaFe$_{0.9}$Mn$_{0.1}$AsO$_{1-y}$F$_y$ ($y = 0.1 - 0.5$) contained small amounts of LaAs (< 3 wt %, for $y = 0.1$ and $0.2$) and with increasing $y$ additional Na$_{1.5}$La$_{1.5}$F$_6$ ($y = 0.3, 0.4, 0.5$) and LaF$_2$ ($y = 0.4, 0.5$) were identified as impurities, with the highest amount (sum of 20 *wt* %) for the highest nominal fluorine content. Since the effective F content cannot be derived from Rietveld refinements or EDX measurements, the evolution of the lattice parameters was used as indicator of the effective F content in the samples. As one can see from Fig. 1 (right side), additional F in LaFe$_{0.9}$Mn$_{0.1}$AsO leads to a decrease of the lattice parameter *a* while the *c* axis is more or less unaffected after a small decrease for $y = 0.1$. Since the sample quality is very good for low F concentrations, the effective F content should be very similar to the nominal one in the low doping regime. The decreasing cell volume indicates that the effective F content further increases with nominal *y*, despite the higher amount of impurity phases. We therefore show the nominal fluorine contents, keeping in mind that higher *y* also means higher effective F content (even though they might not be linearly related). Together with the shrinkage of the *a*-axis also the Fe-Fe bond length is decreased, which reaches a value similar to optimally electron doped LaFeAsO$_{0.9}$F$_{0.1}$, for $y = 0.3$. The two-fold As–Fe–As angle $\varepsilon_2$ further decreases with increasing *y* and almost reaches the ideal value of 109.47° for $y = 0.5 (\approx 110°)$. Because of the shrinkage of the *a*-axis (together with the $\varepsilon_2$ angle) and the nearly unaffected length of the *c* axis, a geometrical consequence is a further increase of the metal-arsenic distance. From a structural point of view, with additional F doping in $LaFe_{0.9}Mn_{0.1}AsO$ the precondition for superconductivity appears to be fulfilled. We find metal-metal distances similar to optimally electron doped LaFeAsO and an As–Fe–As angle close to the ideal value. Counterproductive to this is the increasing metal-arsenic distance caused by the initially increase of the *c* axis by Mn doping. An increasing bond length can lead to more localized electrons due to a smaller orbital overlap.

A metal to semiconductor transition for Mn concentrations higher than 3 % and an increasing semiconducting behavior with increasing Mn content in $LaFe_{1-x}Mn_xAsO$ was reported.[7] The semiconducting behavior is associated to the suppression of the structural transition, which was concluded from thermopower measurements.[7] Our findings are consistent also for further Mn doping as seen from Fig. 2. With increasing *x* the semiconducting behavior gets more pronounced, and the ratio $R_{12K}/R_{300K}$ is the highest for $x = 0.2$. The measured resistivity $\rho_s$ at 300 K increases as well and it is more than one order of magnitude higher for the highest Mn concentration, compared to undoped LaFeAsO ($4.7 \cdot 10^{-4}$ Ωm for $x = 0.2$ and $2.8 \cdot 10^{-5}$ Ωm for $x = 0.0$). Furthermore low temperature XRD measurements showed that the structural transition is absent for the whole series $LaFe_{1-x}Mn_xAsO$ ($x = 0.05-0.20$). Additional electrons introduced by F doping lead to a higher charge carrier concentration within the FeAs-layer and the metallic behavior is regained with increasing F doping. Figure 2 shows that F doping in $LaFe_{0.9}Mn_{0.1}AsO_{1-y}F_y$ leads to a decrease of the $R_{12K} / R_{300K}$ values and the resistivity $\rho_s$ at 300 K. The discordant values for $\rho_{s(300K)}$ of the $y = 0.3$ and 0.5 sample are probably due to increased contact resistances. We did not find superconductivity for the formally electron doped samples ($y > 0.2$) nor did we observe an anomaly of the electrical resistivity for charge compensated $LaFe_{0.9}Mn_{0.1}AsO_{0.9}F_{0.1}$. Low temperature XRD measurements of $LaFe_{0.9}Mn_{0.1}AsO_{1-y}F_y$ ($y = 0.10$ and $0.20$) showed no transition from tetragonal to orthorhombic symmetry at low temperatures.

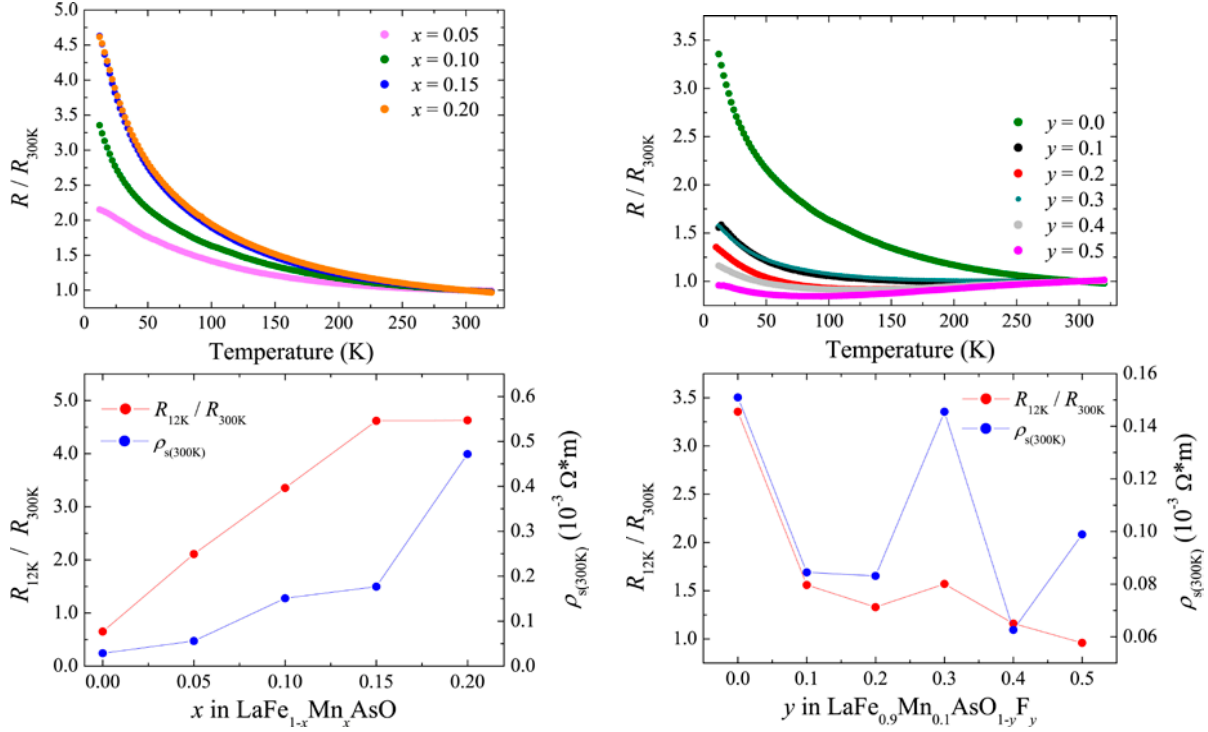

FIG. 2. Temperature dependence of the normalized resistivity $R/R_{300K}$ (top) and evolution of $R_{12K}/R_{300K}$ and resistivity $\rho_{s(300K)}$ (bottom) for $LaFe_{1-x}Mn_xAsO$ (left side) and $LaFe_{0.9}Mn_{0.1}AsO_{1-y}F_y$ (right side).

Magnetic measurements on a SQUID magnetometer showed temperature independent susceptibility and no anomaly that would indicate a magnetic transition. In some samples ferromagnetic impurities were found. Under the assumption that this impurity is metallic iron, we estimated the amount to be smaller than 0.5 *wt* %.

For a more detailed study of the magnetic properties we performed muon spin relaxation (μSR) experiments on the $x$ = 0.05-0.20 and the $y$ = 0.10 and 0.20 samples. μSR as a local magnetic probe can provide valuable information on the magnetic volume fraction and the magnetic homogeneity. Figure 3 shows the zero field (ZF) μSR spectra for the Mn doped samples ($y$ = 0). The data of LaFeAsO ($x$ = 0) are shown for comparison.[27] At high temperatures the muon spin polarization is only weakly relaxing as a function of time due to the interaction of the muon spin ensemble with the small magnetic fields originating from nuclear magnetic moments or diluted ferromagnetic impurities only. At low temperatures anyhow the muons might experience a much stronger internal magnetic field due to ordering of the electronic moments. This is the case for all Mn doped samples as evident from the strongly time dependent muon spin polarization observed in the ZF spectra. In a long range ordered magnet a coherent muon precession of the whole ensemble is observed giving rise to long-lived oscillations in the ZF μSR time spectra as it is the case for the $x$ = 0.0 sample. The value of the precession frequency is proportional to local magnetic field and therefore to the ordered electronic magnetic moment. A damping of the μSR oscillation indicates a distribution of internal magnetic fields sensed by the muon ensemble and is therefore a measure of the disorder in the magnetic system. It is evident that the μSR precession is strongly damped for all Mn- doped samples. This proves that the doping of Mn ions into the magnetic Fe lattice introduces considerable disorder making the magnetic ordering short range in nature. The magnetic correlation length can be estimated with a rule of thumb: If the precession is just visible as in the $x = 0.05$ sample the magnetic correlation length is about 10 lattice constants only.[28] Interestingly the observed frequency is still 16.5 MHz for the $x = 0.05$ sample compared to 23 MHz at $x = 0$. If the same stripe AFM magnetic structure is assumed this means that the average ordered magnetic moment is only reduced by 29% for this doping level. It should be noted that it seems unlikely that the stripe AFM order is realized since no structural distortion could be observed for $x > 0.05$. From our local probe (μSR) data alone it is not possible to deduce the magnetic structure. Therefore it is also not possible to decide if the apparent magnetic disorder stems from localized magnetic Mn ions within a disordered stripe AFM phase or if it is due to a disordered mixture of different antiferromagnetic phases as e.g. observed in Cr doped $Ba(Fe_{1-x}Cr_x)_2As_2$ [9] or due to a new disordered magnetic structure e.g. of Néel-type as observed in LaMnAsO or $BaMn_2As_2$.[29, 30] To clarify this point magnetic neutron scattering data would be indispensable.

With a local probe like ZF μSR on the other hand it is possible to determine the magnetic volume fraction. In a 100% static magnetically ordered powder, 2/3 of the internal field components are perpendicular to the initial muon spin direction and cause a precession (or fast relaxation) while the remaining 1/3 fraction does not precess. It is clear from figure 3 that in all Mn doped samples the full volume is statically magnetic at 5 K. In a dynamic magnetic state also the remaining 1/3 component would show a relaxation [28] which is not the case here. In magnetically ordered $Ba(Fe_{0.925}Mn_{0.075})_2As_2$ inelastic neutron scattering have detected magnetic spin fluctuations at two different wave vectors corresponding to the stripe and Néel type of magnetic order.[12] Here, we find no indications for spin fluctuations in $LaFe_{1-x}Mn_xAsO$. This does not necessarily mean the absence of these fluctuations, but that the fluctuations, if present, are too fast to be observed within the time window of the μSR technique.

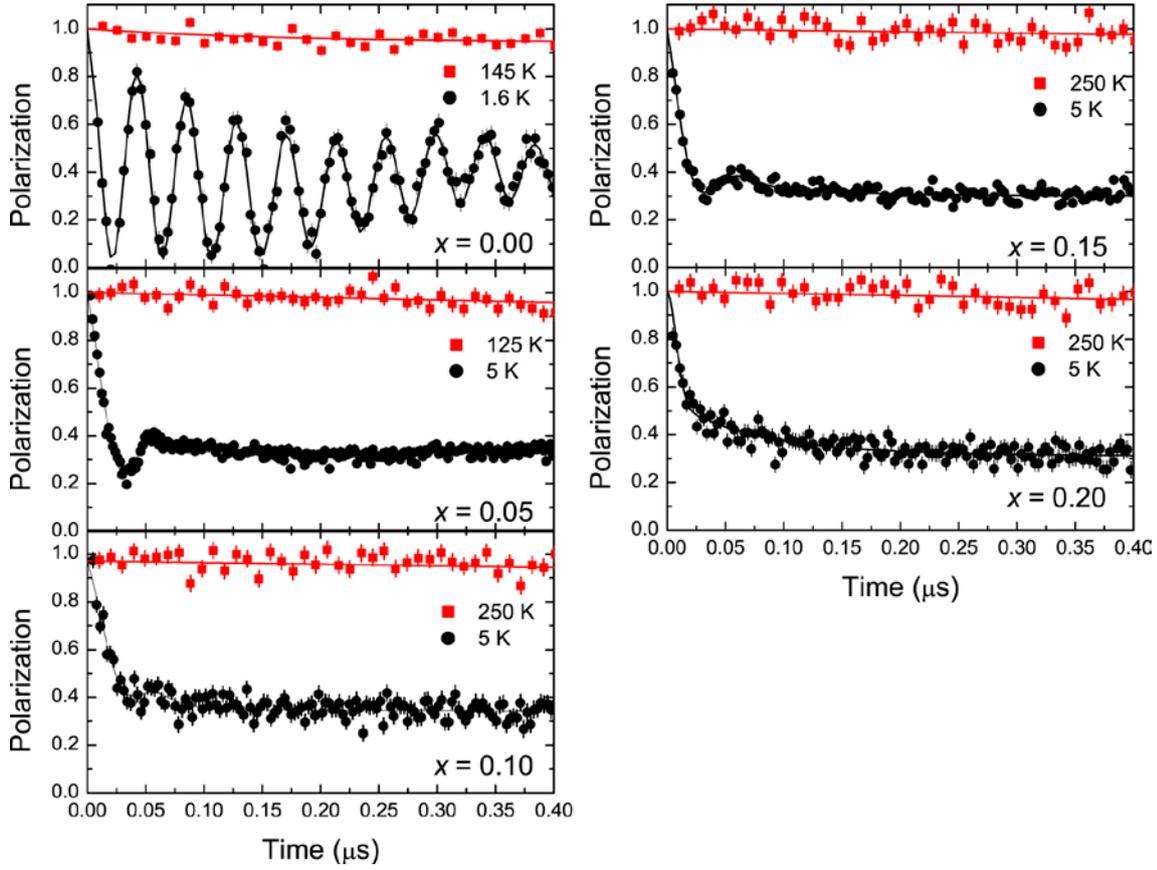

FIG. 3. Zero field µSR spectra of the LaFe$_{1-x}$Mn$_x$AsO series, the data for $x = 0.0$ was taken from [27].

The temperature dependence of the magnetic volume fraction can be better determined in a series of weak transverse field (TF) µSR measurements. In this case a weak external magnetic field of 50 Oe is applied perpendicular to the initial muon spin direction. In a paramagnetic state all the muon spins precess in the external field. If the sample, on the other hand, is magnetic the muon spins precess in the much larger internal fields. Therefore the amplitude of the precession signal in the external magnetic field is a measure for the paramagnetic volume fraction. In Fig. 4 the magnetic volume fraction is shown as a function of temperature for various Mn doping levels. The transition is rather broad for the Mn-doped samples, while it is sharp for the undoped compound. This indicates a certain amount of disorder and/or chemical inhomogeneity. In such a situation it is difficult to determine the true Néel temperature $T_N$, therefore the values for 10%, 50% and 90% magnetic ordering are highlighted.

In Fig. 5 the obtained magnetic phase diagram is shown. It resembles that for Mn doped BaFe$_2$As$_2$ with a minimum of $T_N$ around $x = 0.10$. For Mn concentrations of x ≥ 0.10 in Ba(Fe$_{1-x}$Mn$_x$)$_2$As$_2$ only long-range magnetic order has been detected so far.[10, 11] On the contrary our data indicate short range / disordered magnetism for the 1111 system which can be understood if Mn acts predominately as a magnetic scattering centre, which presumably distorts the long range magnetic order of the iron atoms, or induces a different type of short range antiferromagnetic order itself. Additionally, it is natural to assume that the SDW magnetism of the Fe sublattice which is observed for the parent compound is destroyed or at least fairly disturbed by Mn doping with x ≥ 0.05 since the magnetoelastically coupled orthorhombic distortion is absent for these samples.

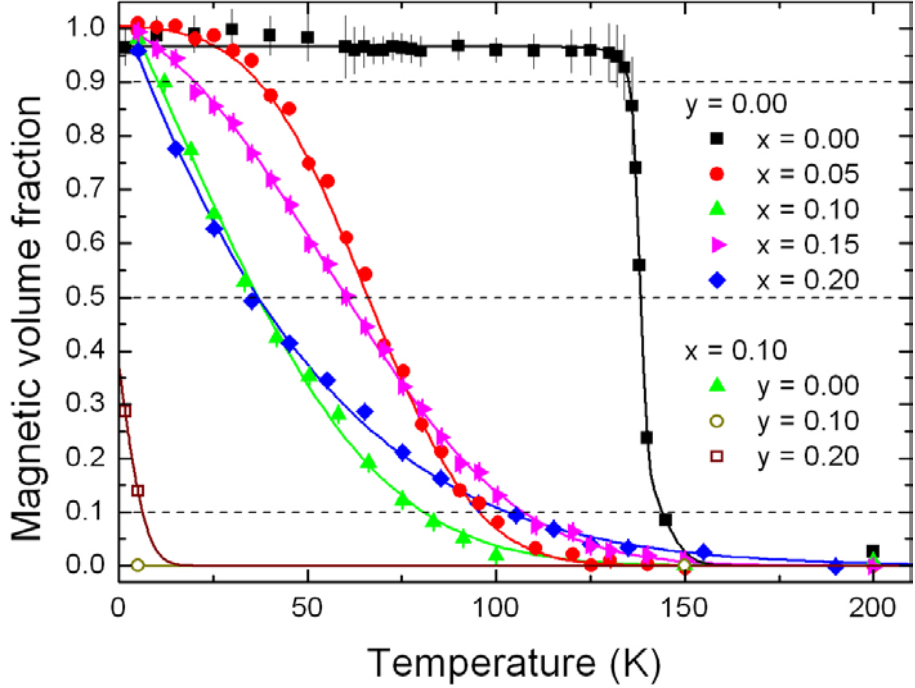

FIG. 4. Development of the magnetic volume fraction as a function of temperature for $LaFe_{1-x}Mn_xAsO_{1-y}F_y$, data for $x = 0.0$ are taken from Ref. 27.

The ZF spectra of the charge compensated ($y = 0.10$) and formally electron doped sample ($y = 0.20$) are shown in Fig. 6 (together with $LaFe_{0.9}Mn_{0.1}AsO$). Surprisingly, F doping leads to a complete suppression of the magnetic transition and the $y = 0.1$ sample is nonmagnetic over the whole temperature range. The observation of an essentially non-magnetic state for a charge compensated sample is astonishing since it is completely different from $Ba_{1-x}K_x(Fe_{0.93}Co_{0.07})_2As_2$ ($x \approx 0.14$), where the magnetism of the parent compound is regained for the charge compensated composition.[21] Actually we would like to point out that in the case of the 1111 family investigated here the charge compensated sample with $x = y = 0.1$ is the most non-magnetic sample investigated in this study.

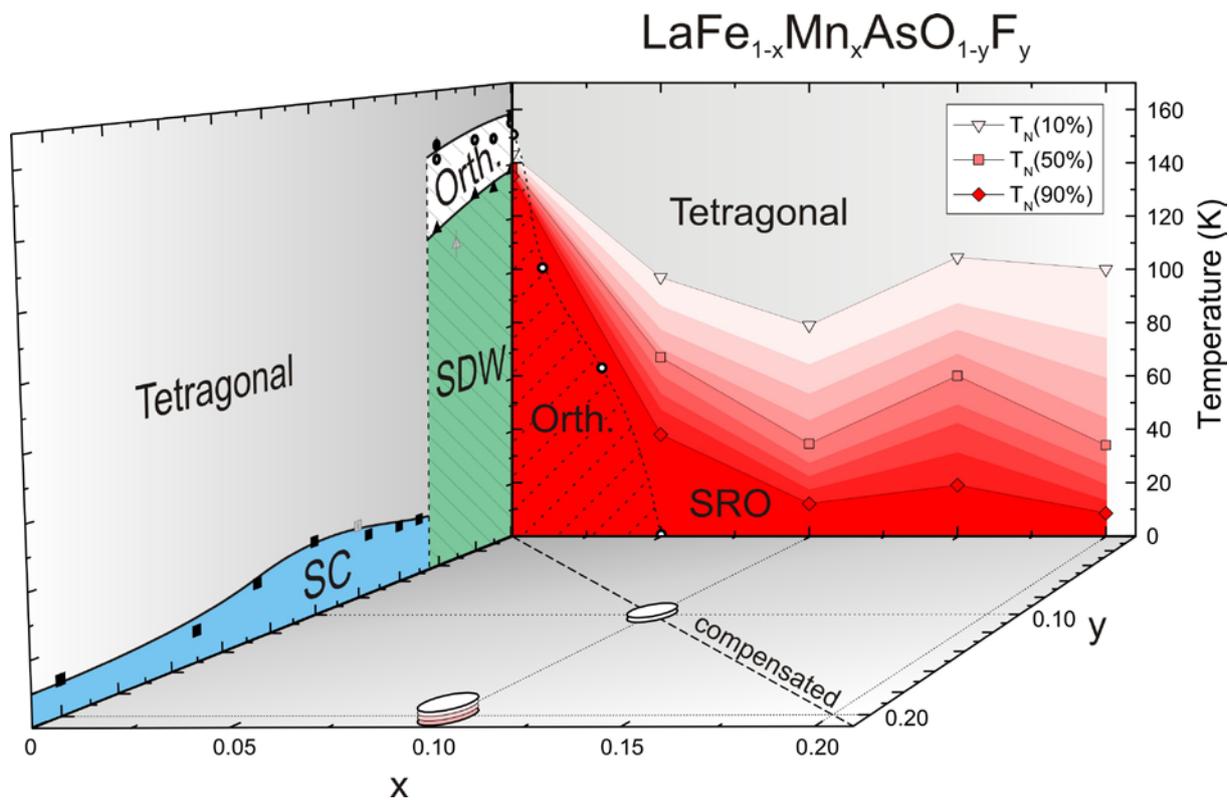

FIG. 5. Structural and magnetic phase diagram of LaFe$_{1-x}$Mn$_x$AsO$_{1-y}$F$_y$. The structural data for $x < 0.05$ are taken from Ref. 7. The data for $x = 0.0$ are taken from Ref. 27. The magnetic volume fraction has a color code in steps of 10% from white (0% magnetic) to red (90% magnetic). The compensated sample with $x = 0.1$ and $y = 0.1$ is non-magnetic down to the lowest measured temperature while weak magnetism develops in 30% of the sample in the $x = 0.1$, $y = 0.2$ sample below 5 K.

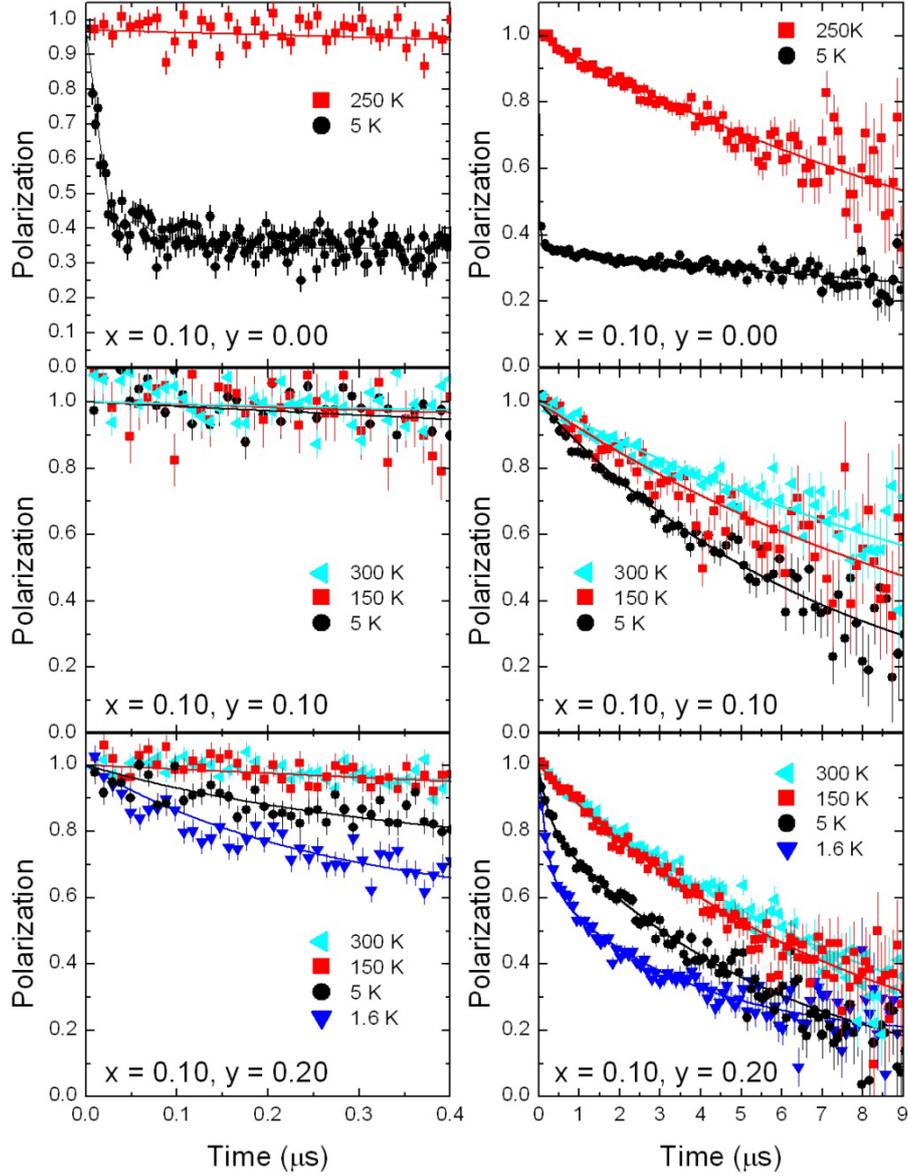

FIG. 6. Zero field spectra of $LaFe_{0.9}Mn_{0.1}AsO$ and $LaFe_{0.9}Mn_{0.1}AsO_{1-y}F_y$ ($y = 0.1$ and $0.2$). On the left side a short time window is shown as for the Mn doped series presented in Fig. 3 and on the right a long time window is displayed, since the magnetism is much weaker for the F doped compounds.

The $y = 0.2$ sample is similar to the $y = 0.1$ sample, but the magnetism is a slightly stronger which can be appreciated by comparing the 5 K data in Fig. 6. ZF, TF and LF µSR measurements show that the observed weak magnetism remains partially dynamic down to the lowest measured temperatures and that it does only occupy ≈ 30% of the sample volume. In none of the investigated samples superconductivity could be observed. This means that although the magnetism can be suppressed or strongly weakened with further F doping it is not possible to induce superconductivity in F doped $LaFe_{0.9}Mn_{0.1}AsO$.

From a NMR study Texier *et al.*[6] found that introducing Mn into $Ba(Fe_{1-x}Mn_x)_2As_2$ does actually not introduce charge doping, and that the Mn ion carries a local moment due to the localization of the additional hole. These localized moments in turn can couple to the Fe electronic band and induce an alternating spin polarization into it. For the case of Ba(Fe$_{1-x}$Mn$_x$)$_2$As$_2$ it was argued that these local Mn moments are unable to suppress the antiferromagnetic ordering at low doping, but that they suppress superconductivity due to breaking of Cooper pairs. Texier *et al.* [6] speculate that this pair breaking should prevent superconductivity even if the long range magnetic order could be destroyed by other means. Similar conclusions can be drawn from our structural, magnetic and transport measurements on LaFe$_{1-x}$Mn$_x$AsO$_{1-y}$F$_y$. The doping with relatively small amounts of $x \geq 0.05$ Mn into LaFe$_{1-x}$Mn$_x$AsO leads to a complete suppression of the tetragonal to orthorhombic transition usually concomitant with the stripe-like SDW AFM order. Upon further Mn doping short range magnetism is found to persist up to the highest doping level studied here. It is possible that a similar but more disordered kind of Néel magnetic order as observed for Cr and Mn doped BaFe$_2$As$_2$ compounds is established, even though we cannot directly prove this hypothesis with our local probe technique. As in the case of Ba(Fe$_{1-x}$Mn$_x$)$_2$As$_2$ no superconductivity is induced by Mn doping into LaFeAsO. On the contrary, the room temperature resistivity increases with Mn doping and the temperature dependence of the resistivity changes from metallic to increasingly semiconducting and the increasing bond lengths indicate a more localized electronic behavior. The electron doping by introducing F into of LaFe$_{0.9}$Mn$_{0.1}$AsO$_{1-y}$F$_y$ in contrast increases the conductivity of the system as evidenced by our resistivity measurements. In addition, the static magnetism is quickly suppressed by the electron doping and the structural prerequisites for high-$T_c$ superconductivity like an almost regular Fe-As tetrahedron is successively approached. Nevertheless neither in the nominally charge compensated compound ($x = 0.1$, $y = 0.1$) nor in the nominally electron doped compound ($x = 0.1$, $y = 0.2$) superconductivity is induced. Theoretically it has been shown, that the combination of short-range Néel fluctuations and pair-breaking impurity scattering effectively can suppress superconductivity.[16] Therefore it is reasonable to assume that a similar effect is at work in LaFe$_{0.9}$Mn$_{0.1}$AsO$_{1-y}$F$_y$ as well with localized paramagnetic Mn magnetic moments and possibly residual Néel fluctuations acting as pair breakers in the otherwise non-magnetic samples.

## IV. SUMMARY

We obtained high quality samples of LaFe$_{1-x}$Mn$_x$AsO using a solid state metathesis reaction. Also the double substituted LaFe$_{0.9}$Mn$_{0.1}$AsO$_{1-y}$F$_y$ series has been obtained in very good quality in a one-step reaction, despite the use of 5 starting materials. Structural investigations revealed that upon additional F doping, parameters like the metal-metal distance or the As–Fe–As angle reach values which have been thought to be essential for the emergence of high superconducting $T_c$s. On the other hand, the increase of the metal-arsenic distances indicates a situation with stronger localization of the electrons. The magnetic behavior of the Mn doped 1111 compounds is different from the corresponding 122 compounds. We find short range magnetic order, with the transition temperatures passing a minimum for $x = 0.10$. While the structural transition is present for Mn concentrations of up to $\approx 11\%$ in $Ba(Fe_{1-x}Mn_x)_2As_2$, the $LaFe_{1-x}Mn_xAsO$ series with $x \geq 0.05$ shows no structural transition down to 10 K. The previously reported semiconducting character of $LaFe_{1-x}Mn_xAsO$ ($x = 0$-$0.1$) [7] was confirmed and gets more pronounced with higher Mn concentrations. Together with the semiconducting character, the measured $\rho_{s(300K)}$ values are further increased with increasing Mn concentration. Additional electron doping with F leads in turn to a more metallic behavior of the resistivity. The magnetic transition is completely suppressed for charge compensated $LaFe_{0.9}Mn_{0.1}AsO_{0.9}F_{0.1}$, which means that charge compensation does not lead to the regain of the parent compound's magnetic properties as in charge compensated $Ba_{1-x}K_x(Fe_{0.93}Co_{0.07})_2As_2$ ($x \approx 0.14$).[21] The nominal optimal electron doped $LaFe_{0.9}Mn_{0.1}AsO_{0.8}F_{0.2}$ shows very weak magnetism in only 30% of the sample at temperatures below 5 K only. Thus although the magnetic transition is suppressed and the electron count should lead to a superconducting state, the latter is not observed down to low temperatures. As stated above, it has been proposed that in such a case the localized Mn moments and possible residual Néel fluctuations may act as pair breakers. In agreement with this theoretical prediction, our results show that Mn impurities within the FeAs layer are detrimental to superconductivity in electron doped $LaFeAsO_{1-y}F_y$. In summary, the series $LaFe_{1-x}Mn_xAsO_{1-y}F_y$ shows a complex structural, electronic and magnetic phase diagram in which electron and hole doping have very different electronic and magnetic effects. This is in stark contrast to the $Ba_{1-x}K_x(Fe_{1-y}Co_y)_2As_2$ system where the electron count essentially governs the physical properties.[21]

## ACKNOWLEDGEMENT

This work has been supported by FP7 European project SUPER-IRON (grant agreement No.283204).